\documentclass[10pt,journal,compsoc]{IEEEtran}
\ifCLASSOPTIONcompsoc

  \usepackage[nocompress]{cite}
\else
  \usepackage{cite}
\fi

\usepackage[hidelinks]{hyperref}

\usepackage[cmex10]{amsmath}
\usepackage{array}
\usepackage{url}
\usepackage{amssymb}
\usepackage{color}
\usepackage{stfloats}
\usepackage{tabularx}
\usepackage[table]{xcolor}
\usepackage{booktabs}
\usepackage{multirow,graphicx}
\usepackage{tcolorbox}
\usepackage{colortbl}
\tcbuselibrary{skins}
\usepackage{blindtext}
\usepackage{floatrow}
\usepackage{algorithm}
\usepackage{algpseudocode}
\usepackage{float}
\floatstyle{plaintop}
\restylefloat{table}
\usepackage{chngcntr}
\usepackage[catalan,english]{babel}
\usepackage{soul}

\begin{document}
%
\title{Fast Quasi-Conformal Regional Flattening of the Left Atrium}
%
%
%
%

\author{{Marta~Nu\~nez-Garcia\hyperlink{https://orcid.org/0000-0001-9349-3059}{\includegraphics[scale=0.08]{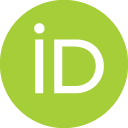}}
Gabriel~Bernardino\hyperlink{https://orcid.org/0000-0001-8741-2566}{\includegraphics[scale=0.08]{figs/ORCIDiD_icon128x128.png}},
Francisco~Alarc\'on\hyperlink{https://orcid.org/0000-0002-5713-9551}{\includegraphics[scale=0.08]{figs/ORCIDiD_icon128x128.png}},
Gala~Caixal\hyperlink{https://orcid.org/0000-0003-1830-3125}{\includegraphics[scale=0.08]{figs/ORCIDiD_icon128x128.png}}, 
Llu\'is~Mont\hyperlink{https://orcid.org/0000-0002-8115-5906}{\includegraphics[scale=0.08]{figs/ORCIDiD_icon128x128.png}}, 
Oscar~Camara\hyperlink{https://orcid.org/0000-0002-5125-6132}{\includegraphics[scale=0.08]{figs/ORCIDiD_icon128x128.png}} 
and~Constantine~Butakoff\hyperlink{https://orcid.org/0000-0002-8526-5045}{\includegraphics[scale=0.08]{figs/ORCIDiD_icon128x128.png}}}
        

\IEEEcompsocitemizethanks{\IEEEcompsocthanksitem M. Nu\~nez-Garcia, G. Bernardino, O. Camara and C. Butakoff are with Physense, Department of Information and Communication Technologies,
Universitat Pompeu Fabra, Barcelona, Spain.\protect\\
E-mail: marta.nunez@upf.edu

\IEEEcompsocthanksitem F. Alarc\'on, G. Caixal and Ll. Mont are with Department of Cardiology, Unitat de Fibri\lgem aci\'o Auricular (UFA), Hospital Clinic, Universitat de Barcelona and Institut d'Investigacions Biom\`ediques August Pi i Sunyer (IDIBAPS), Barcelona, Spain.}
}

\IEEEtitleabstractindextext{%
\begin{abstract}
Two-dimensional representation of 3D anatomical structures is a simple and intuitive way for analysing patient information across populations and image modalities. While cardiac ventricles, especially the left ventricle, have an established standard representation (bull's eye plot), the 2D depiction of the left atrium (LA) remains challenging due to its sub-structural complexity including the pulmonary veins (PV) and the left atrial appendage (LAA). Quasi-conformal flattening techniques, successfully applied to cardiac ventricles, require additional constraints in the case of the LA to place the PV and LAA in the same geometrical 2D location for different cases. Some registration-based methods have been proposed but surface registration is time-consuming and prone to errors when the geometries are very different. We propose a novel atrial flattening methodology where a 2D standardised map of the LA is obtained quickly and without errors related to registration. The LA is divided into 5 regions which are then mapped to their analogue two-dimensional regions. 67 human left atria from magnetic resonance images (MRI) were studied to derive a population-based template representing the averaged relative locations of the PVs and LAA. The clinical application of our methodology is illustrated on different use cases including the integration of MRI and electroanatomical data.
\end{abstract}

}

\maketitle

\IEEEdisplaynontitleabstractindextext

%
\IEEEpeerreviewmaketitle

\IEEEraisesectionheading{\section{Introduction}\label{sec:introduction}}

%
%
%
%
\IEEEPARstart{M}{EDICAL} imaging techniques are increasingly used nowadays to extract crucial anatomical and functional information used in diagnosis and treatment planing. The imaged organs are inherently three-dimensional and appropriate visualization tools are needed, including manual rotations and different points of view, to get a complete overview of the whole organ under study.

Flattening methods aim to compute an unfolded representation of a 3D surface mesh by projecting it to a simpler 2D domain, easier to visualize, manage and interpret.
Additionally, if the 2D map is standardised (e.g. same anatomical regions of different subjects spatially coincide) the 2D unfolded domains can be used as a common reference space to analyse multi-modal data from different patients or from the same patient at different time-steps. The reader is referred to \cite{kreiser2018survey} for a thorough review of flattening methods applied to human organs, including the brain, different bones, and the vascular system among others.

In the case of the heart, the 17 segment AHA bull's eye plot of the left ventricle (LV) has been widely used by clinicians for long time \cite{cerqueira2002standardized,soto2016integration,paun2017patient}. The LV's conical shape and the absence of salient sub-structures in 3D anatomies derived from medical images (e.g. ignoring the trabeculations) highly facilitates its flattening. In general, only one basal boundary is considered, which is mapped to the outer contour of a 2D disk, and one reference point (the apex) which is placed in the centre of the circle.
On the contrary, unfolding of the LA is challenging since its main cavity is connected to several pulmonary veins, the left atrial appendage and the left ventricle through the mitral valve (MV). Furthermore, there exists a notable variability in the anatomical configuration and spatial location of these sub-structures. 
For instance, the most common LA morphology (up to 70\%) involves 4 PV (with variable size, shape, position and orientation with respect to the main cavity) with occasional presence of a common left trunk, extra right PV or other oddities \cite{prasanna2014variations}. Strong variation can also be seen in the LAA shape and the MV size, shape and position with respect to the PVs \cite{garcia2018sensitivity}. Proposing a common reference space for the LA, including the multiple possible interfaces between the main cavity and its sub-structures, is therefore non trivial.

Ma et al. \cite{ma2012cardiac} were the first researchers proposing a LA flattening technique, based on a B-spline with proportional distance between any pair of points in the 3D LA surface and their mapped pairs on the 2D map. As a result, the different 2D maps were not standardised since the position of the PV holes were not in the same position for different patients.
Karim et al. \cite{karim2014surface} proposed to flatten the LA to a square also without constraining the position of the PV and LAA holes. The authors carefully studied the distortion induced by the unfolding process and showed how the LA flattened square could be used to display and qualitatively analyse different types of data from the same patient. 
However, a direct comparison across different patients was not feasible because of the lack of correspondence between the different maps (e.g. PV holes were not positioned in the same place). 
More recently, Williams et al. \cite{williams2017standardized} proposed a standardised unfold map (SUM) suitable to compare different LA using as reference a 3D LA template and its corresponding 2D flattened version, both related with a defined 3D-2D point-to-point mapping. Their flattening strategy was based on registering an arbitrary LA to the 3D template and then transferring the data to the 2D LA template with the known 3D-2D template mapping. The original LA template was divided into 24 regions, allowing regional analysis of the mapped data. The main limitation of this approach is the difficulty of obtaining an accurate registration between different LA surface meshes due to their high shape variability. Together with errors induced by required data projection and interpolation steps, this scheme leads to undesired information loss between the 3D and 2D LA representations if the LA under study is substantially different from the LA template. Additionally, advanced surface mesh registration techniques such as currents-based ones are associated to large computational times. Finally, a universal atrial coordinate system was developed in \cite{roney2018} and used to represent the left and right atria in a two-dimensional domain. The proposed representation is however not completely standardised since the position of the veins and appendages may be distinct for the different cases.

In this paper, we propose a quasi-conformal \cite{levy2002least,lui2013} (in general, there is no conformal map compatible with a given map along the boundary) flattening of the LA with the following boundary constraints: the MV contour is mapped to the external circumference of a 2D disk, and the PVs and LAA ostia contours are mapped to predefined circumferences within the disk. Quasi-conformal flattening of surface meshes with holes often results in undesired mesh self-folding (holes appear covered by adjacent mesh cells) and to overcome this issue, we impose additional regional constraints. Five anatomical regions are defined in the 3D LA which are afterwards confined to their 2D counterparts. Our method is fast (almost real-time) and without information loss (all points in the 3D mesh are represented in the 2D domain).
In order to define a realistic two-dimensional template, the relative position of the PVs and LAA as well as the different PV and LAA ostium sizes need to be considered. We computed the average position of these sub-structures analysing a population of 67 atrial shapes and used them to define an intuitive and realistic unfolded representation of the LA, that at the same time minimizes the distortion inherent to any flattening technique. To illustrate the proposed methodology we used several synthetic datasets, real LA from Late-Gadolinium Enhanced Cardiac Magnetic Resonance (LGE-CMR) data and real electroanatomical maps.

\section{Methods}
\begin{figure*}[th]
\centering
\includegraphics[width=0.94\textwidth]{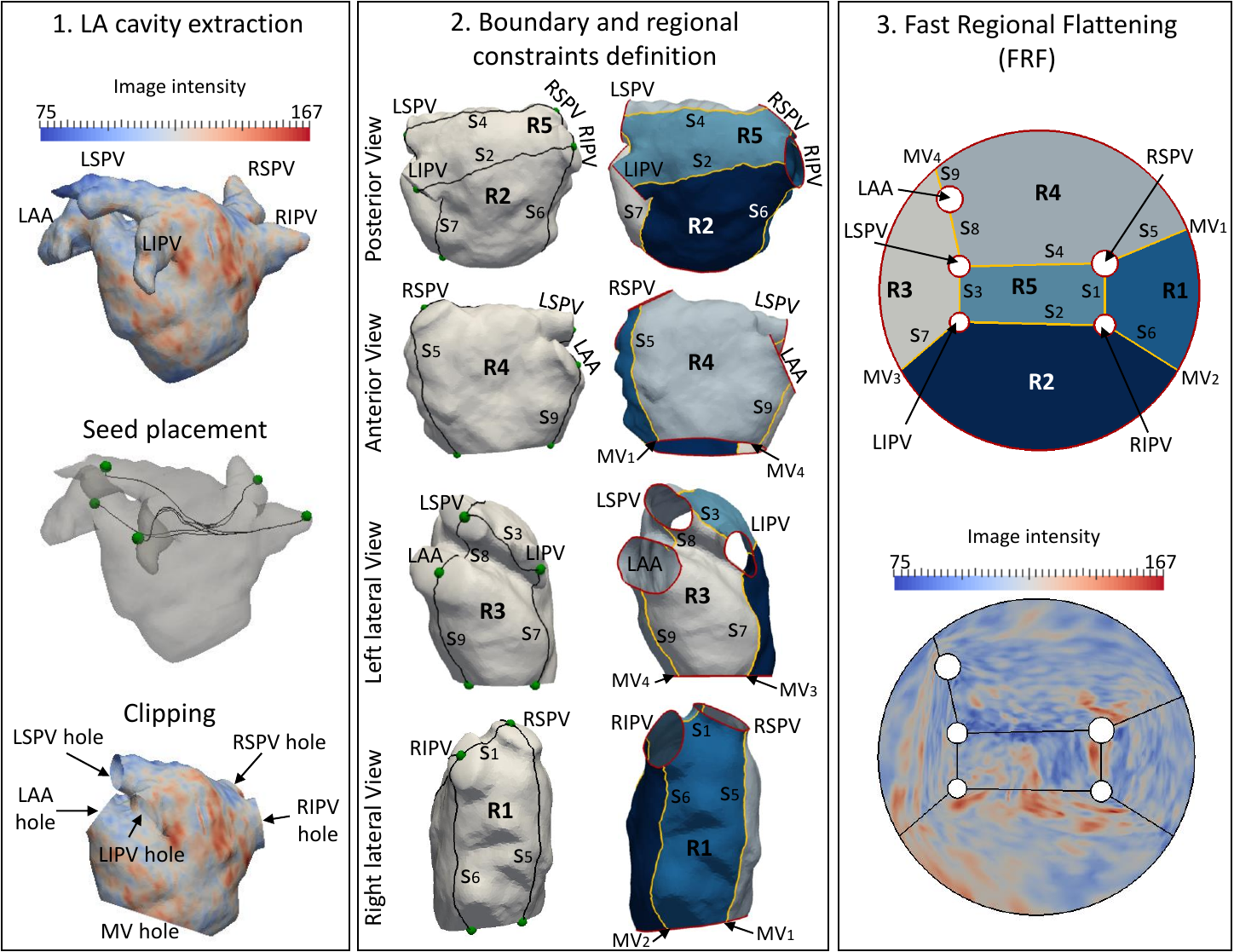} 
\caption[Overview of proposed flattening method]{Overview of the proposed method: (1) LA cavity extraction; (2) boundary and regional constraints definition (LA division); (3) Fast Regional Flattening (FRF). Six boundaries (displayed in red) corresponding to the MV, 4 PV, and LAA ostia are used. Nine segment paths (s$_{1-9}$, in yellow) are used to define 5 LA regions (R1-R5). Data mapped into the LA surface mesh correspond to intensity values from the associated late-gadolinium magnetic resonance image (blue and red colors indicate low and high image intensity, respectively). LSPV = left superior PV; LIPV = left inferior PV; RSPV = right superior PV; RIPV = right inferior PV; LAA = left atrial appendage; MV = mitral valve; MV$_{1}$, MV$_{2}$ = seed points in the MV contour delimiting the septal wall, and MV$_{3}$, MV$_{4}$  = seed points in the MV contour delimiting the left lateral wall.}
\label{fig:pipeline}
\end{figure*}

A scheme of the complete flattening framework can be seen in Fig. \ref{fig:pipeline} and it is described in section \ref{sec:flat_pipeline}. The methodology employed to define the most appropriate 2D template is detailed in section \ref{subsec:template_definition}.

\begin{figure*}[ht!]
\centering
\includegraphics[width=\textwidth]{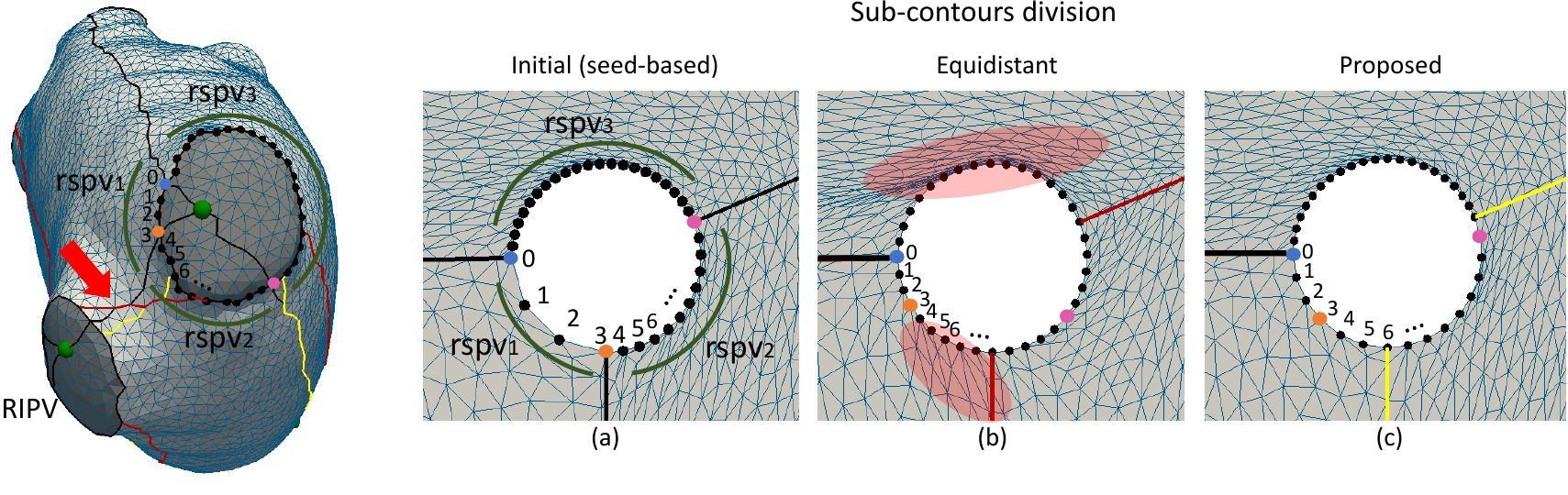}
\caption[Constraints recomputation.]{Constraints recomputation. On the left, 3D LA surface mesh (RSPV detail) with different divisions considered: initial seed-based division (black paths); division achieved imposing equidistant points in the hole (red paths); and our proposed division (yellow paths). Contour points (vertices) are enumerated, and initial sub-contour extremes (intersection points) are shown in blue (fixed reference point), orange and pink. On the right, flattened RSPV ostium contour using the original geodesic paths (a), recomputed paths using equidistant points (b) and recomputed paths using the proposed division (c). The RSPV point seed (green) is slightly displaced to the border of the RSPV ostium to exaggerate this effect in the example. rspv$_{i}$ = $i$-th sub-contour of RSPV ostium contour; RSPV = right superior PV.}
\label{fig:subcontours}
\end{figure*}

\subsection{Flattening pipeline}
\label{sec:flat_pipeline}

The method comprises three main steps: (1) LA cavity extraction; (2) boundary and regional constraints definition; and (3) Fast Regional Flattening (FRF).

\subsubsection{LA cavity extraction}
\label{subsec:step1}
Given any arbitrary 3D surface mesh representing a LA, the first step aims at standardising the LA shape by only keeping its main cavity after semi-automatically cutting the PVs, the LAA and the MV. This cutting process requires to manually place 5 seeds near the ending points of the PVs and the LAA. The reader is referred to \cite{tobon2015benchmark} for more details on this method.

\subsubsection{Boundary and regional constraints definition}
\label{subsec:step2}
In the second stage, holes in the main LA cavity corresponding to interfaces with the PV and the LAA are closed and triangulated using the method described in \cite{liepa2003filling}. Then, 9 seed points are manually placed in specific and easily recognisable atrial regions: 5 at the centre of the closed PV and LAA ostia, and other 4 on the MV contour: 2 delimiting the interatrial septal wall and 2 delimiting the left lateral wall. 
Inter-seed geodesic paths (i.e. shortest curve between two points on a mesh, such that the curve lies on the surface \cite{mitchell1987discrete}, s$_{1}$-s$_{9}$ in Fig. \ref{fig:pipeline}) are computed using the Dijkstra shortest path algorithm \cite{Dijkstra1959} dividing the LA surface into 5 anatomical regions (R1-R5 in Fig. \ref{fig:pipeline}):

\begin{itemize}
\item R1 defines the inter-atrial septal wall, which is delimited by inter-seed geodesic connecting the 2 right PVs (s$_{1}$), geodesics connecting them with the MV septal seeds (s$_{5}$ and s$_{6}$), and the corresponding piece of MV contour between MV$_{1}$ and MV$_{2}$.
\item R2 corresponds to the part of the posterior wall and LA floor delimited by the 2 inferior PVs (s$_{2}$), the geodesics connecting them with the MV contour (s$_{6}$ and s$_{7}$), and the MV contour between MV$_{2}$ and MV$_{3}$.
\item R3 corresponds to the left lateral wall and it is delimited by the geodesics connecting the MV contour with the LAA (s$_{9}$), the LAA with the LSPV (s$_{8}$), the two left PVs (s$_{3}$), the LIPV with the MV contour (s$_{7}$), and the MV contour between MV$_{3}$ and MV$_{4}$.
\item R4 defines the anterior wall, which is bounded by geodesics connecting the MV contour and the RSPV (s$_{5}$), both superior veins (s$_{4}$), the LSPV and LAA (s$_{8}$), the LAA and MV contour (s$_{9}$), and the MV contour between MV$_{4}$ and MV$_{1}$.
\item R5 corresponds to the part of the posterior wall framed by the geodesics connecting the 4 PVs (s$_{1-4}$).

\end{itemize}

After that, the final LA cavity that will be flattened is obtained by automatically removing the PV and LAA hole covers. The \emph{boundary constrained points} are identified as the PV, LAA and MV boundary points (mesh vertices of the red curves in Fig. \ref{fig:pipeline}), and the \emph{regional constrained points} are identified as the projection of the inter-seed paths (s$_{1}$-s$_{9}$) onto the final LA cavity (points of the yellow curves in Fig. \ref{fig:pipeline}).

In our framework, the PV and LAA ostium contours are mapped to discretised circumferences and their correct representation depends on the sampling (total number of vertices or points) of the ostia contours: a small number of points induces circumferences with large edges, while a high number of points produces circumferences with very short edges.
The LAA and PV ostia contours are actually divided into 2 and 3 sub-contours, respectively, as a consequence of the LA regional division explained above, further influencing how the circumferences are represented in the 2D domain. 
Initially, these sub-contours are determined by the inter-seeds paths which likely cause non-equidistant point distribution in the contour (Fig. \ref{fig:subcontours}a). An equidistant distribution of points (Fig. \ref{fig:subcontours}b) can be achieved by conveniently displacing the sub-contour point extremes (or intersection points: blue, orange and pink dots in Fig. \ref{fig:subcontours}), but this could potentially worsen the anatomical meaning of the LA division (note the red arrow showing how the red line connecting RSPV and RIPV is twisted), and induce mesh distortion (see areas with stretched triangles (highlighted with semi-transparent red ellipses in Fig. \ref{fig:subcontours}b). 
Defining as \textit{proportional length} the number of points of each sub-contour corresponding to an equidistant distribution of points in the complete hole, and \textit{real length} as the number of points of the sub-contour derived from the initial LA division, a good compromise between suitable distribution of points and meaningful LA division is achieved by imposing a sub-contour length equal to $ \textit{real length} +  \left \lfloor (\textit{proportional length} - \textit{real length})/2  \right \rfloor$ number of points. Note that the second addend can be either positive or negative. One intersection point is fixed (point with index 0 and blue color in Fig. \ref{fig:subcontours}) and the remaining two are updated using the previous rule. This modifies the number of points in each sub-contour but the total number of points remains the same. Using the updated sub-contour extremes, the \emph{boundary} and \emph{regional constrained points} are recomputed.  

\subsubsection{Fast Regional Flattening (FRF)}
\label{subsec:step3}

\begin{figure*}[th]
\centering
\includegraphics[width=0.85\textwidth]{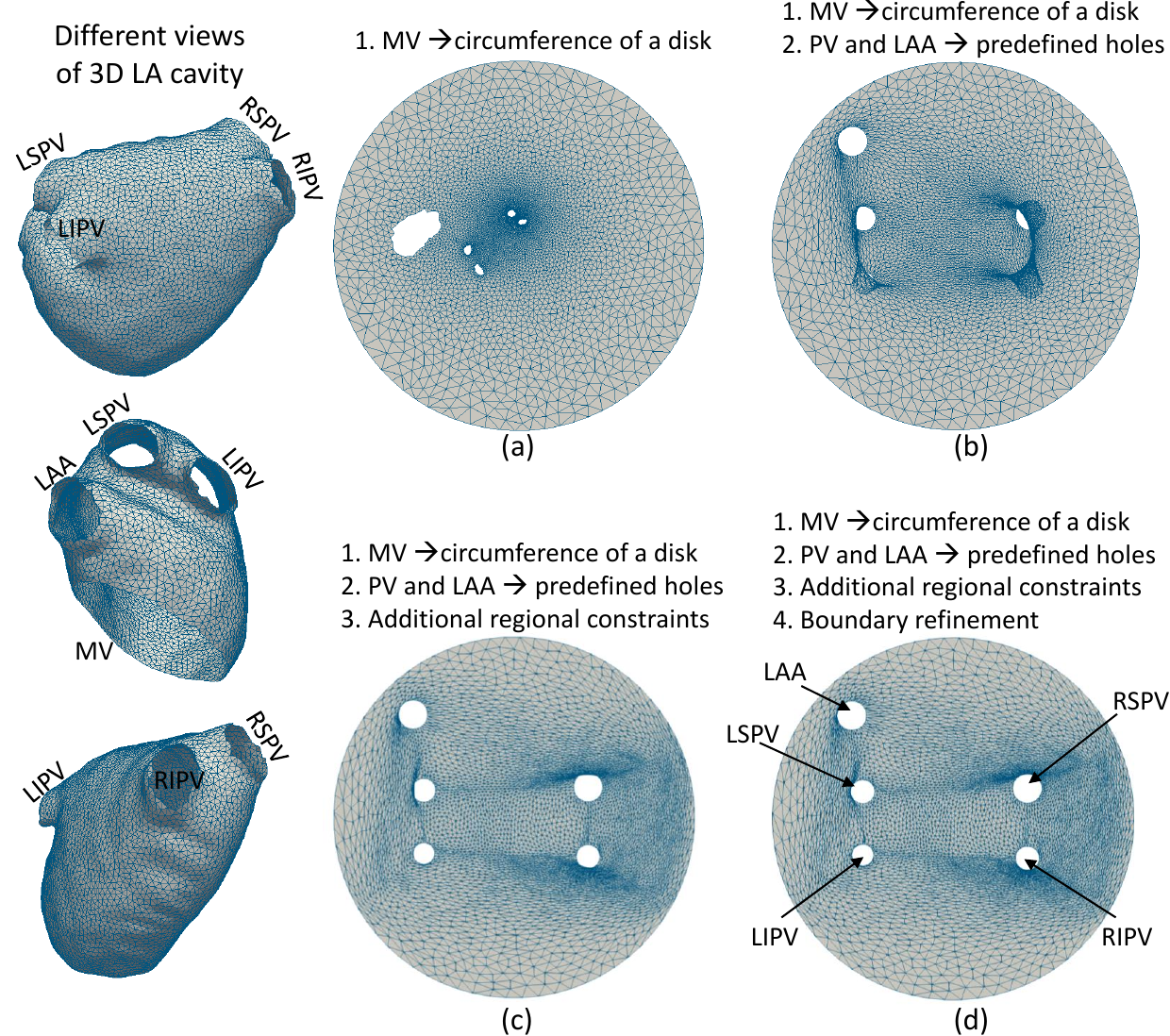} 
\caption[Example of different quasi-conformal LA flattenings.]{Example of 3D LA surface mesh (left column, 3 different views) and different LA unfolding strategies: (a) quasi-conformal flattening constraining only the MV to the external boundary of a 2D disk; (b) quasi-conformal flattening additionally fitting the PV and LAA hole contours; (c) quasi-conformal flattening fitting PV, LAA contours and imposing regional constraints; (d) final flattening after boundary refinement of (c) mesh. LSPV = left superior PV; LIPV = left inferior PV; RSPV = right superior PV; RIPV = right inferior PV; LAA = left atrial appendage; MV = mitral valve.}
\label{fig:meshes_flat}
\end{figure*}

\subsubsection*{Overview}
We aim to obtain a quasi-conformal (i.e. angle-preserving) and standardised flat representation of the 3D LA cavity surface mesh where the boundaries (MV, PV and LAA holes) are constrained to predefined circumferences within the 2D disk. An example of conformal LA flattening where only the MV boundary position is imposed (i.e. only MV boundary is mapped to the circumference of the 2D disk) can be seen in Fig. \ref{fig:meshes_flat}a. All holes except the MV were closed prior to the flattening and opened afterwards.
In order to additionally restrict the PV and LAA ostia contours to predefined positions we add the corresponding boundary constraints to the quasi-conformal scheme. By doing that, the boundary points appear correctly placed in the targeted positions but the holes are often covered by adjacent triangles since the method does not avoid mesh self-folding (Fig. \ref{fig:meshes_flat}b).
To overcome this issue, we include additional regional constraints in the parameterisation, but minor triangle overlapping can still occur near the holes (Fig. \ref{fig:meshes_flat}c). To further refine the boundary we recompute the point coordinates on the 2D flattened LA with a quasi-conformal parameterisation only constrained with the boundary points and not the regional constraints (Fig. \ref{fig:meshes_flat}d).

\subsubsection*{Mathematical framework}

Let $M$ be the LA cavity surface mesh with N vertices (points) and $M_{F}$ its corresponding flattened mesh. 
Let $I_{\partial M}$ be the indices of the \emph{boundary points} (MV, PV, LAA) of $M$. Let $\Delta_{M}$ be the $N\times N$ (cotangent \cite{pinkall1993computing}) Laplacian of $M$ and let $\Delta_{M}^{'}$ be $\Delta_{M}$ where the off-diagonal elements of the rows defined by the positions of $I_{\partial M}$ set to 0 and corresponding elements on the main diagonal set to 1.
Let \textbf{$\left(b_{x},b_{y}\right)$} be the coordinates of the 2D \emph{boundary points} of $M_{F}$ in the same order as the corresponding indices appear in $I_{\partial M}$ and let $\boldsymbol{b}_{x}^{'}$ and $\boldsymbol{b}_{y}^{'}$ be N-dimensional vectors with values $\boldsymbol{b}_{x}$ and $\boldsymbol{b}_{y}$, respectively, in the positions of $I_{\partial M}$ and zeros elsewhere. 

Let P be the number of \emph{regional constrained points} (i.e. number of points in the dividing segments (s$_{1}$-s$_{9}$), yellow lines in Fig. \ref{fig:pipeline}) and $I_{s}$ the corresponding point (vertex) indices. Let $\left(s_{x},s_{y}\right)$ be the 2D coordinates of the P vertices in the same order as they appear in $I_{s}$ and let $\boldsymbol{E}_{s}$ be a $P \times N$ zero matrix with 1 in each row in the positions corresponding to vertices of $I_{s}$ (i-th row has 1 in the position given by the i-th element of $I_{s}$).

In order to find the coordinates $\left(\boldsymbol{x}^{*},\boldsymbol{y}^{*}\right)$ of the vertices of $M_{F}$, we propose to solve the following two quadratic programming problems: 

\begin{equation}
\boldsymbol{x}^{*}=\arg\min_{\boldsymbol{x}} \left(w \left\Vert \boldsymbol{\Delta}_{M}^{'}\boldsymbol{x}-\boldsymbol{b}_{x}^{'}\right\Vert ^{2}\right) \hspace{0.5cm}
\textnormal{s.t.} \hspace{0.5cm} \boldsymbol{E}_{s}\boldsymbol{x}=\boldsymbol{s}_{x}
\end{equation}

\begin{equation}
\boldsymbol{y}^{*}=\arg\min_{\boldsymbol{y}} \left(w \left\Vert \boldsymbol{\Delta}_{M}^{'}\boldsymbol{y}-\boldsymbol{b}_{y}^{'}\right\Vert ^{2}\right) \hspace{0.5cm}
\textnormal{s.t.} \hspace{0.5cm} \boldsymbol{E}_{s}\boldsymbol{y}=\boldsymbol{s}_{y}
\end{equation}

\noindent
with $w$ (set to 1000 in our experiments) penalizing the unfulfilment of the \emph{boundary constraints}. Using Lagrange multipliers these can be rewritten and solved as a system of linear equations:

\begin{align}
\left[\begin{array}{cc}
w^{2}\boldsymbol{\Delta}_{M}^{'T}\boldsymbol{\Delta}_{M}^{'} & \boldsymbol{E}_{s}^{T}\\
\boldsymbol{E}_{s} & \boldsymbol{0}
\end{array}\right]\left[\begin{array}{c}
\boldsymbol{x}^{*}\\
\boldsymbol{\lambda}_{x}
\end{array}\right]&=\left[\begin{array}{c}
w^{2}\boldsymbol{\Delta}_{M}^{'T}\boldsymbol{b}_{x}^{'}\\
\boldsymbol{s}_{x}
\end{array}\right]\\
\left[\begin{array}{cc}
w^{2}\boldsymbol{\Delta}_{M}^{'T}\boldsymbol{\Delta}_{M}^{'} & \boldsymbol{E}_{s}^{T}\\
\boldsymbol{E}_{s} & \boldsymbol{0}
\end{array}\right]\left[\begin{array}{c}
\boldsymbol{y}^{*}\\
\boldsymbol{\lambda}_{y}
\end{array}\right]&=\left[\begin{array}{c}
w^{2}\boldsymbol{\Delta}_{M}^{'T}\boldsymbol{b}_{y}^{'}\\
\boldsymbol{s}_{y}
\end{array}\right]
\end{align}

\noindent

To refine the boundary we proceed as follows. Let $\boldsymbol{\Delta}_{F}$ be the $N\times N$ Laplacian of $M_{F}$ and let $\Delta_{F}^{'}$ be $\Delta_{F}$ 
where the off-diagonal elements of the rows defined by the positions of $I_{\partial M}$ set to 0 and corresponding elements on the main diagonal set to 1.
The refined coordinates $\left(\boldsymbol{x}^{*'},\boldsymbol{y}^{*'}\right)$ of the vertices of $M_{F}$ in the final flat representation can be found by solving the following system of linear equations:

\begin{align}
\boldsymbol{\Delta_{F}^{'}}\boldsymbol{x}^{*'} &= \boldsymbol{b}_{x}^{'}\\ 
\boldsymbol{\Delta_{F}^{'}}\boldsymbol{y}^{*'} &= \boldsymbol{b}_{y}^{'}
\end{align}

\begin{figure}[t]
\centering
\includegraphics[width=\columnwidth]{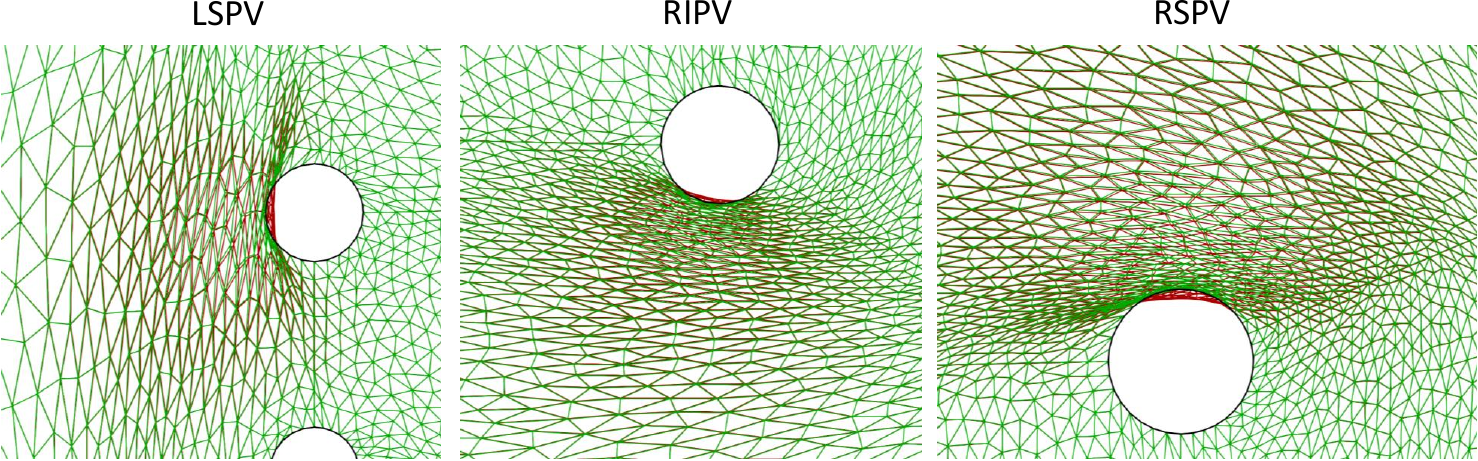}
\caption[Zoom of PV holes region after flattening]{Zoom of PV holes region after the initial flattening (red) and after boundary refinement (green). Target boundaries are shown in black. LSPV = left superior PV; RIPV = right inferior PV; RSPV right superior PV.}
\label{fig:puntos}
\end{figure}

\noindent
The need for this boundary refinement is illustrated in Fig. \ref{fig:puntos} where the initial flattening is depicted in red and the final solution is depicted in green. 

Pseudocode corresponding to the complete flattening procedure is shown in Algorithm \ref{flat_algorithm}.

\begin{algorithm}[t]
\caption{Complete flattening pipeline} \label{flat_algorithm}
\begin{algorithmic}[1]
\Procedure{Flat}{$mesh$}\Comment{triangular mesh}
\State Clip PV, LAA, MV 
\State Cover PV, LAA holes
\State Divide LA: select and connect 9 seeds. 
\State Remove PV, LAA covers $\rightarrow$ get $M$, \emph{boundary} and \emph{regional constrained points}. Get \emph{intersection points} (\emph{IP}) as \emph{boundary} $\cap$ \emph{regional constrained points}. Get all PV \emph{sub-contours}: portion of a \emph{boundary} between 2 consecutive \emph{IP}. Each PV \emph{boundary} is made by the concatenation of 3 \emph{sub-contours}     
\State Recompute constraints: \For{each PV \emph{boundary}}
\begin{enumerate}
\item Compute L$_{12}$, L$_{23}$, L$_{31}$, \emph{real length} (i.e. number of points) of each \emph{sub-contour}
\item Compute P$_{12}$, P$_{23}$, P$_{31}$, \emph{proportional length} of each \emph{sub-contour}
\item Fix \emph{IP}$_{1}$
\item Displace \emph{IP}$_{2}$, $\left \lfloor \frac{P_{12}-L_{12}}{2}  \right \rfloor$ points 
\item Displace \emph{IP}$_{3}$, $\left \lfloor \frac{P_{23}-L_{23}}{2} \right \rfloor$ points 
\end{enumerate}
\EndFor      

\State Using all new \emph{IP}, update \emph{boundary} and \emph{regional constrained points}
\State Flatten $M$ by solving eq. (3-4)
\State Refine the solution by solving eq. (5-6)

\State \textbf{return} $M_{F}^{'}$ 
\EndProcedure
\end{algorithmic}
\end{algorithm}

\subsection{2D template definition}
\label{subsec:template_definition}

\subsubsection*{Overview}

To define the size and position of the PV and LAA circumferences so that they reflect their physiological relative location as much as possible, we analysed 67 manually segmented LA (image resolution of 0.625 x 0.625 x 1.25 mm$^{3}$) from the 2018 Atrial Segmentation Challenge\footnote{http://atriaseg2018.cardiacatlas.org/} characterised by 4 separate PVs. The remaining atria from the challenge (33) presented different number of PVs or merged LAA and LSPV due to segmentation errors or imaging limitations. We first homogenised the meshes by clipping the PVs, LAA and MV using the procedure proposed by \cite{tobon2015benchmark}.
We measured the perimeter of PVs and LAA ostia; left and right carina lengths (i.e. distance between ipsilateral (same side) veins (s$_{1}$ and s$_{3}$ in Fig. \ref{fig:pipeline})); distance between the 2 inferior and the 2 superior veins, respectively (s$_{2}$ and s$_{4}$); distance between LSPV and LAA (s$_{5}$); and distance between right and left veins and the MV (s$_{6}$-s$_{9}$). We also used the 67 shapes to build a mean shape or template (shown in Fig. \ref{fig:meshes_flat}) using Deformetrica \cite{bone2018deformetrica}.

Besides considering a realistic relative position of PVs and LAA, we additionally aimed to reduce distortion in the disk and achieve intuitive visualization. The small MV contour is mapped to a fairly bigger region, i.e. the contour of the disk, while at the same time constraining the remaining surface to the interior of the disk. This fact naturally induces stretching of the outer part of the disk and shrinkage of the interior (see Fig. \ref{fig:meshes_flat}a) causing high compression of information in the centre. Because of that, we created two additional templates modifying the measured PV and LAA relative positions to enlarge the central part of the disk and achieve better visualization. 

\begin{figure*}[th]
\includegraphics[width=\textwidth]{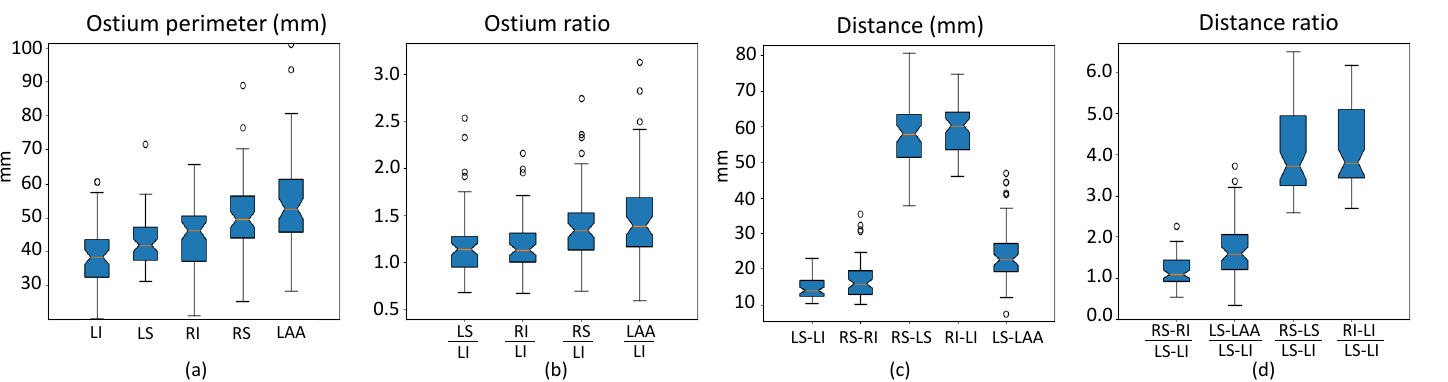} 
\caption[Summary of shape measurements]{Summary of shape measurements. (a) PV and LAA ostium perimeters; (b) ratio between remaining ostium perimeters and LIPV perimeter that is the smallest one and considered as reference; (c) inter PV and LAA distances; (d) ratio of the inter PV and LAA distances and the LSPV-LIPV separation (left carina) which is considered as reference. All distances are reported in mm. LI = left inferior PV; LS = left superior PV; RI = right inferior PV; RS = right superior PV; LAA = left atrial appendage.}
\label{fig:boxplots}
\end{figure*}

An ideal 3D-2D mapping should transform the mesh elements (i.e. triangles) in an isotropic way and preserving their area. Therefore, to analyse the quality of the different LA unfolded maps, we defined distortion indices based on area preservation and transformation isotropy. 
To quantify the area preservation we first normalised the size of the 3D and the corresponding 2D LA surfaces to have areas equal to 1. Then, we computed for each triangle the ratio between its area in the 2D disk and in the 3D surface (let $\boldsymbol{\alpha}$ be the ratio). As the total areas were previously normalised, having $\boldsymbol{\alpha}$ bigger than 1 implies triangle enlargement, smaller than 1 implies compression and equal to 1 reflects perfect size preservation. Constant ratio in the whole mesh is desired since it means that all triangles are equally transformed in terms of size.    
To inspect the isotropy of the transformation we used the Jacobian matrix of the 3D-2D transformation \cite{levy2002least}:

\begin{equation}
\boldsymbol{J} = \begin{pmatrix}
\partial u/\partial x &  \partial u/\partial y\\ 
 \partial v/\partial x &  \partial v/\partial y
\end{pmatrix}
\end{equation}
where $(x,y)$ are point coordinates in the local orthonormal coordinate system of each triangle and $(u,v)$ are the global coordinates of the flat domain.
The two eigenvalues of $\boldsymbol{J}$ characterise the deformation of the triangle with the isotropy estimated as the ratio between the smallest and largest eigenvalue (let $\boldsymbol{\beta}$ be the ratio, $\boldsymbol{\beta} = $  1 meaning isotropic transformation, and decreasing $\boldsymbol{\beta}$ meaning anisotropic triangle stretching). 

We additionally considered the Spatial Context Preservation (SCP), defined in \cite{kreiser2018survey} as a qualitative index that helps the user to contextualize the observed structures: the better the spatial context is preserved, the easier a certain region can be visually related to its corresponding position in the original 3D data. 
To inspect the SCP of the different templates we created a synthetic texture pattern of same size circles uniformly distributed on the 3D surface mesh and inspected how the circles were displayed in the different templates.

\subsubsection*{Template (FRF map) creation}

\begin{figure*}[th]
\includegraphics[width=\textwidth]{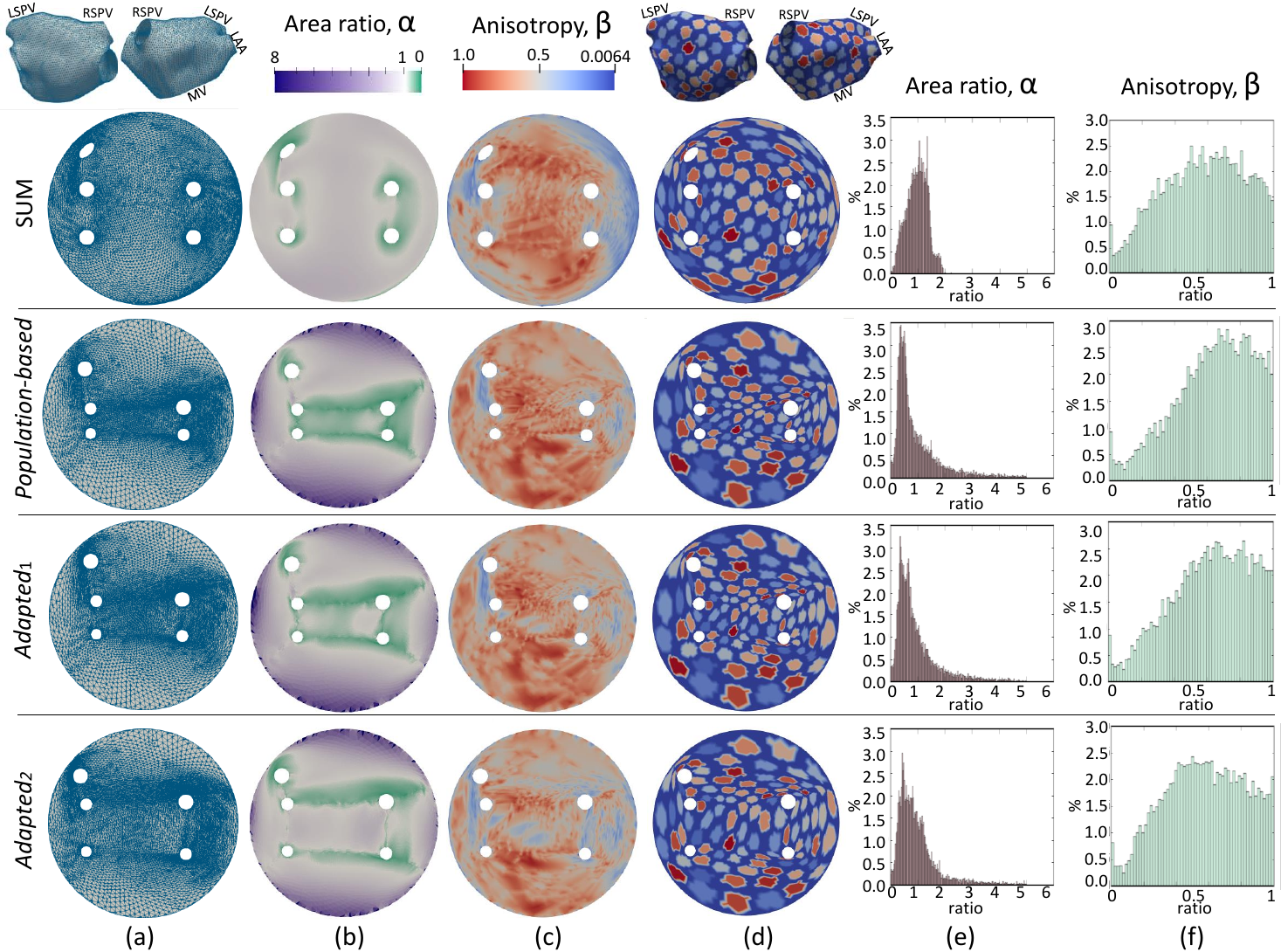} 
\caption[Distortion analysis performed using the 3D SUM template]{Distortion analysis performed using the 3D SUM template. The first row corresponds to the SUM template, and the other 3 to different FRF maps with different hole positions, \emph{population-based} (second row), \emph{Adapted$_{1}$} (third row) and \emph{Adapted$_{2}$} (fourth row). (a) displays the mesh cells (triangles), (b) shows the area-related distortion, (c) the anisotropic distortion metric, (d) the result of mapping uniform circles, (e) histogram of 2D/3D areas ratio ($\alpha$), and (f) histogram of eigenvalues ratio ($\beta$).}
\label{fig:distortion}
\end{figure*}

Measurements corresponding to the PV and LAA ostium perimeters can be seen in Fig. \ref{fig:boxplots}a and b, and in the Appendix (Table \ref{tab:perimeters}).
According to our results the left PVs have significantly smaller perimeter than the right PVs ($p<10^{-6}$). 
In particular, perimeters were related as follows: LIPV $\boldsymbol{<^{*}}$ LSPV $<$ RIPV $\boldsymbol{<^{*}}$ RSPV $<$ LAA where $\boldsymbol{<^{*}}$ denotes significantly smaller at the 5\% significance level.  
Taking as a reference the smallest perimeter (LIPV) we established the following perimeter proportional values: LSPV/LIPV = 1.1; RIPV/LIPV = 1.1; RSPV/LIPV = 1.35; and LAA/LIPV = 1.35, using population median values (Table \ref{tab:perimeters}, right).

Measurements corresponding to the inter-vein distances (s$_{1-4}$) and LSPV-LAA distance (s$_{8}$) are summarized in Fig. \ref{fig:boxplots}c and d, and in the Appendix (Table \ref{tab:distances}). The left carina (i.e. region between the two left veins) was found significantly shorter than the right carina ($p=0.007$) with median ratio of 1.1. 
The separation between LSPV and LAA was found significantly higher that the left carina length ($p<10^{-6}$) with the median proportional ratio of almost 1.6. The separation between the 2 superior and the 2 inferior veins was found comparable ($p=0.37$) and significantly higher than the carinas ($p<10^{-6}$). The proportional ratio was estimated as 3.75 approximately (width of the posterior wall with respect to left carina length). 
Finally, we also found that the distance between left PVs and the MV contour was approximately half of the distance between right PVs and the MV contour.

Once all the ratios were estimated, the remaining step required to define the reference values (i.e. centre and radio of the LIPV circumference, and LIPV-LSPV (left carina) separation). Here the criterion used was to obtain a representation with low and uniform distortion distribution. We considered the SUM proposed by \cite{williams2017standardized} as the state of the art method and compared it with 3 different template (or FRF map) configurations: 
\begin{enumerate}
\item \emph{Population-based}: using the proportions measured in the study described above.
\item \emph{Adapted$_{1}$}: measured proportions are modified to achieve a more detailed representation of the central part of the disk (i.e. inter-veins region). 
\item \emph{Adapted$_{2}$}: measured proportions are further modified to achieve SUM-like hole positions.
\end{enumerate}

\section{Experiments and Results}

\subsection{Distortion analysis of 2D templates}
\label{sub:distortion}

Since the SUM methodology registers any LA to the 3D version of a template and then maps the information to the 2D version of the template, only that mesh (the template) can be fairly compared to our method. Results of flattening the SUM template can be seen in Fig. \ref{fig:distortion} where the first row corresponds to the SUM mapping and the other 3 rows to the different FRF maps tested: \emph{Population-based}, \emph{Adapted$_{1}$}, and \emph{Adapted$_{2}$}. 
The different columns show, from left to right: the flattened mesh; the spatial distribution of the two distortion metrics (triangle's area ratio, $\alpha$, and anisotropy of the transformation, $\beta$); uniform circles mapping; and histograms of the distortion metrics considered. As it can be seen, SUM showed good triangle size preservation (triangles' area ratio close to 1) but with less compact histogram than the FRF maps where even if the difference between enlarged and compressed regions seemed emphasized (b), the histogram showed that the ratio between the areas was more constant than in the SUM case (most of the triangles were consistently reduced and only few triangles were enlarged). The SUM mapping had the worst anisotropic distortion profile (less percentage of values close to 1 (f)), especially close to the outer contour of the disk (c). FRF maps showed less anisotropic distortion specially for \emph{Population-based} and \emph{Adapted$_{1}$} cases.

\begin{figure}[t!]
\centering
\includegraphics[width=\columnwidth]{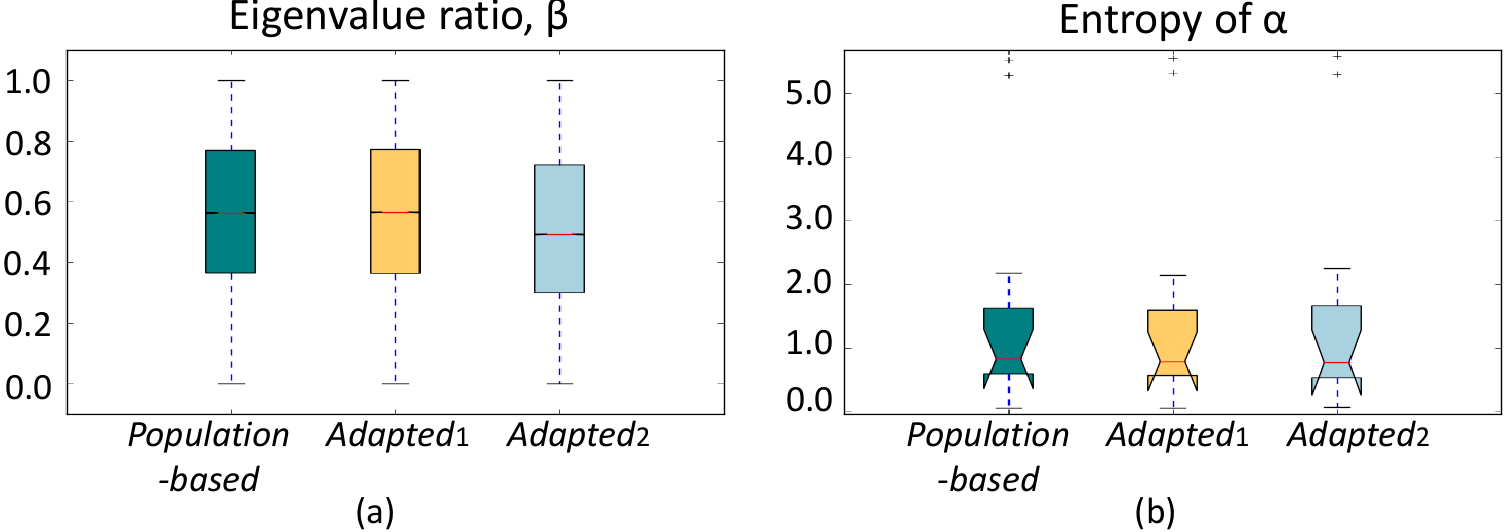}
\caption[Summary of distortion metrics using 15 left atria.]{Distortion metrics in 15 LA. Distortion metrics do not significantly change depending on the templates.}
\label{fig:dist_metrics}
\end{figure}

With regard to the comparison between the different FRF maps using the SUM template, \emph{Population-based} and \emph{Adapted$_{1}$} showed comparable isotropic and area ratio profiles. However, the enlargement of the central part in \emph{Adapted$_{1}$} improves the SCP (see Fig. \ref{fig:distortion}d). \emph{Adapted$_{2}$} was worse with regard to all the considered parameters showing higher anisotropic distortion and worse areas ratio resulting in an excessive and artificial vertical stretching of the central part. 
To decide between the 3 proposed FRF maps, we computed distortion metrics using a cohort of 15 LA. Fig. \ref{fig:dist_metrics}a shows boxplots of the anisotropic distortion metric corresponding to the 15 cases together and, as it can be seen, the three configurations showed comparable results. To determine the template with most compact triangle's size change, we compared the entropy of each histogram. Lower entropy indicates more stable triangle size change which is preferred in this context. As can be seen in Fig. \ref{fig:dist_metrics}b, the distortion is again comparable among the 3 configurations. We chose \emph{Adapted$_{1}$} as the best compromise between isotropic transformation, triangle's area consistency and visual interpretability.

\begin{figure*}[th]
\centering
\includegraphics[width=0.85\textwidth]{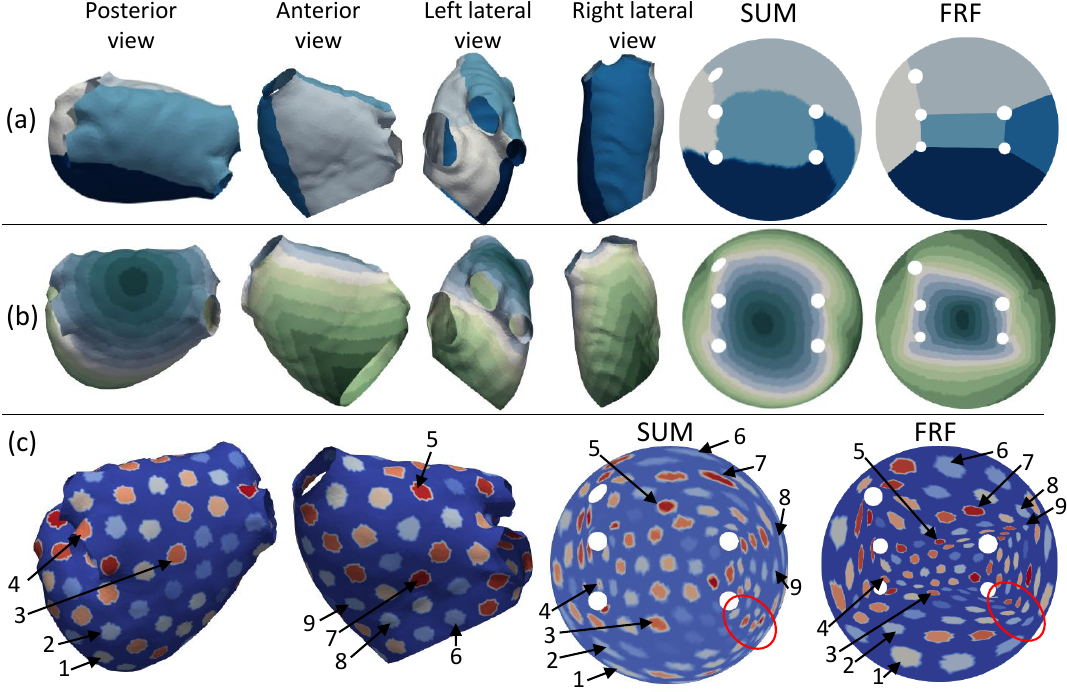} 
\caption[Flattening examples of 3D LA with different synthetic texture patterns on the surface.]{Flattening examples of 3D LA with different synthetic texture patterns on the surface. From top to bottom: 4 views of a sample LA divided into the 5 anatomical regions considered in our approach and regional correspondence to SUM and FRF maps (a); striped synthetic texture, defined on the 3D surface representing isocontours of the distance to a seed point placed in the posterior wall (b); and synthetic texture of 100 uniformly distributed spots (c). Some spots have been numbered in the different representations for clarification purposes.}
\label{fig:circles}
\end{figure*}

The regional relation between the final FRF map chosen (\emph{Adapted$_{1}$}) and the SUM was also studied, and it can be seen in Fig. \ref{fig:circles}a how the same regions are placed in different areas in the two 2D disks. We also used synthetic texture patterns to inspect the nature of the deformations induced by the FRF and compare it to the SUM mapping. Fig. \ref{fig:circles}b shows a striped pattern representing isocontours of the distance to a point placed in the centre of the region between the 4 veins. Fig. \ref{fig:circles}c shows results of mapping a spotted pattern. As it can be seen, circles placed close to the MV appear highly distorted (stretched) in the SUM (e.g. spots number 1, 2, 6 or 8) while the effect of the regional flattening can be observed in the FRF map: note the orientation change of the spots in the region encircled in red, and the different size and shape of spots 8 and 9 (that should be similar since they are close to each other) because they belong to regions R4 and R1, respectively. Importantly, SUM is affected by interpolation errors due to information mapping (in this case, color of the spot) from the 3D LA to the SUM template (note the resolution loss as the circles appear blurred, e.g. in the region encircled in red, and the background color change). On the contrary, using the FRF method, there is no information loss or interpolation errors since all the points in the 3D mesh are represented in the 2D map. Additional synthetic examples can be seen in the Appendix (Fig. \ref{fig:pattern_appen}).

\begin{figure*}[th]
\includegraphics[width=\textwidth]{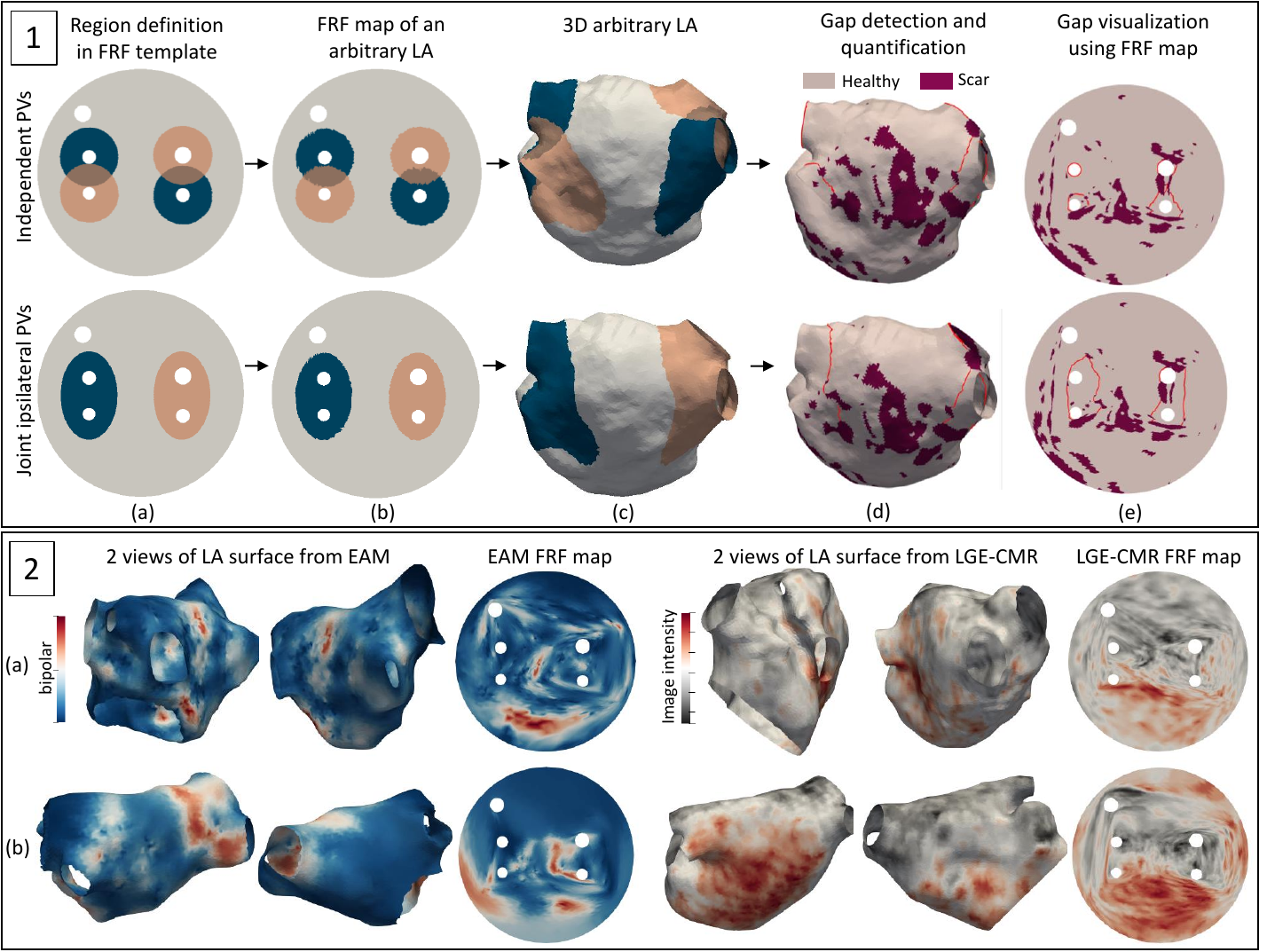} 
\caption[Examples of clinical applications of FRF.]{Examples of clinical applications of FRF. 1. Gap quantification using the FRF method: from left to right: sub-regions are defined in a high resolution FRF map template as the surrounding of each PV (top row) or the surroundings of the two ipsilateral (same-side) PVs (bottom row); the defined subdivision is transferred to the FRF map obtained from an arbitrary LA (b) and then to the corresponding 3D arbitrary LA (c); automatic detection and quantification of gaps (in red) is performed on the 3D LA (d) and optionally displayed on the disk (e). 2. Two examples of electroanatomical maps (EAM) and corresponding (same patient) LA from LGE-CMR with projected image intensity. Low voltage (dark blue) is typically associated to high LGE-CMR intensity (red). LGE-CMR = Late Gadolinium Enhanced Cardiac Magnetic Resonance.}
\label{fig:gaps}
\end{figure*}

\section{Clinical applications}
Several examples of FRF applied to the visualization of LGE-CMR image intensities are shown in the Appendix (Fig. \ref{fig:example}). 
Besides the usefulness of the method for visualization purposes, in the following subsections, we present two clinical applications: quantification of incomplete ablation lines (gaps) after pulmonary vein isolation (PVI) and joint analysis of bipolar voltage information from electroanatomical maps (EAM) and LGE-MRI intensity data. 

\subsection{Gap quantification.} 
PVI is a common procedure for the treatment of AF that aims to electrically isolate the PVs from the LA body producing (with radiofrequency, laser or cryoballoon) a continuous lesion completely encircling the PVs. The resulting lesion is often incomplete being a combination of scar and gaps of healthy tissue that are potential causes for AF recurrence. We have previously developed a semi-automatic method to locate and quantify ablation gaps once a parcellation of the LA is available \cite{nunez2019mind}. 
Different parcellations can be defined in our FRF map and used to consistently divide different 3D LA in order to apply the gap quantification method. An example of the procedure can be seen in Fig. \ref{fig:gaps}-1. To identify gaps after PVI, the surroundings of the different PVs independently (Fig. \ref{fig:gaps}-1, top row) or in pairs (Fig. \ref{fig:gaps}-1 bottom row) must be inspected. The regions where the gaps are sought for were first defined in a high resolution FRF map used as template and were then transferred to the arbitrary FRF map using a closest point mapping and to the 3D LA using the mesh point identifier. After that, the automatic detection and quantification of the gaps was performed as described in \cite{nunez2019mind}.

\subsection{Inspection of the relation between EAM and LGE-CMR data}

The proposed FRF maps can also be used to inspect multi-modal data from the same patient. The relation between voltage in EAM and signal intensity in LGE-CMR data has been investigated providing contradictory results \cite{spragg2012initial,bisbal2014cmr,harrison2015repeat}. Fig. \ref{fig:gaps}-2 shows two examples of EAM invasively acquired during an electrophysiology study (using CARTO (Biosense Webster Inc, Diamond Bar, CA) mapping system) and corresponding (i.e. same patient) LA segmented from LGE-CMR with image intensity projected onto the surface. More examples are shown in the Appendix (Fig. \ref{fig:eam_reso}). It can be seen how FRF puts into correspondence the two different (due to the different acquisition methods) LA anatomies facilitating their qualitative and quantitative comparison. 

\section{Discussion}

The aim of the work presented in this paper is twofold: (1) to develop a method to flatten the LA to a common reference space (i.e. standardised 2D template), and (2) to define this common reference space in a way that provides a clinically meaningful visualization. 

The latter is crucial in our approach and, contrary to methods that focus on minimizing distortion in mathematical sense (e.g. Teichmuller mapping\cite{Lui2014, ng2014}, Ricci flow\cite{wang2011}), we prioritize generating standardised and intuitive 2D depictions of the LA while keeping the distortion low. 
Our approach provides better spatial correspondence between different LA geometries by partitioning the geometry into clinically meaningful regions in 3D and mapping them to their 2D representations preventing vertices from moving outside of their regions (a similar approach can be found in \cite{zhang2014} where 3D registration with exact landmark correspondence is performed by cutting the mesh into different patches and registering them separately). That is not the case of the global parameterisation techniques where region boundaries can get unrealistically displaced. The distortion induced by this regional behaviour of the flattening remains low as shown in Section \ref{sub:distortion}. 

Another important aspect is that in our approach flattening and standardisation is performed simultaneously: when all the conditions are set, the mesh is flattened to a standardised domain in 1 step (2 steps considering the final refinement which is just a small correction). To the best of our knowledge, in most of the publications, the constraints are applied through registration between the 2D domains \cite{wang2011}. Our approach is therefore simpler, requiring less steps, and it can deal with highly variable shapes where registration may fail. In fact, in our experiments, 2D registration (using thin-plate splines, Teichmuller mapping and Currents) between conformally flattened LAs resulted in high distortion due to huge differences in the locations of the PVs and LAA. 
Additionally, if one would use non rigid registration based on diffeomorphic free-form deformation (DFFD), it is typically slow while our method is almost real time after manual seed placement. 


The proposed pipeline can be easily adapted to other chambers of the heart and to other anatomies (typically less complex than the cardiac chambers) such as the ones mentioned in \cite{kreiser2018survey}: the descending aorta, the placenta, vessels, etc. It will require to find a suitable template (disk, square, rectangle, with or without inner holes) and the appropriate constraints. It could also be applied to the flattening of cortical hemispheres where the regionality of our approach may be useful: brain areas such as frontal, temporal, parietal, and occipital lobes could be constrained to specific 2D areas while maintaining, approximately, the conformality of the parameterisation.

We thoroughly compared our method to the SUM mapping, the state of the art method for atrial standardised representation.
Contrary to the SUM method, FRF does not depend on registration being therefore faster and not affected by potential registration errors. Additionally, FRF does not lose any information from the initial 3D surface, i.e. all points are depicted in the 2D map, while SUM needs to interpolate data when mapping the information from the arbitrary 3D surface to the template. This is especially noticeable in the flattening of the spotted patterns (color represents spot identifier) where the texture appears blurred after the flattening process (Fig. \ref{fig:distortion}c). This effect is even more notable in the example in Appendix (Fig. \ref{fig:pattern_appen}c) where it can be seen how many spots have merged due to resolution loss. These errors could be reduced by remeshing the SUM template (increasing the number of points) but this will increase the registration computational cost. The execution time of the registration using the original SUM template (mesh with 7402 points) is around 20 minutes \cite{nunez2019mind} while FRF lasts 2-3 s, after manual seed points placement (Intel i5 3.3 GHz CPU and 16 GB RAM).

The distortion induced by the two methods is different as showed in Fig. \ref{fig:distortion}. SUM method performs a global mapping where the central part is enlarged and the areas close to the MV are pushed and stretched along the contour of the disk. On the other hand, FRF restricts each area to its specific 2D domain (enhancing regional interpretability) even if this fact causes local stretching/shrinkage and distortion close to the limits of the 5 defined regions. 

SUM provides more detailed regional division (24 regions \cite{williams2017standardized} vs. our 5 regions) but this regional division may be inaccurate due to the small size of some regions and the influence of registration errors. FRF brings more control on the 3D regional division by manual placement of seed points and, in the flattening, constraining each region to its specific 2D area. Our algorithm is however highly influenced by this manual seed placement step, especially regarding the seeds placed on the MV contour. The seed points and inter-seed paths define the constraints used in the flattening which is the key part of the method. Automatically computing the constraints, i.e. automatically dividing the LA is challenging due to, for example, bulges in the LA and obliqueness of the MV plane. Nonetheless, since the process of seed placement and unfolding is almost real time it can be repeated several times until a satisfactory result is obtained. 
Additionally, more regions could be defined in our FRF template to further subdivide any arbitrary LA as in Fig. \ref{fig:gaps}-1 a-c.

In our approach, the quality of the mesh impacts the final mapping since inter-seed paths (i.e. \emph{regional constraints}) are computed along the edges of the triangular mesh. To reduce the impact of the specific mesh elements we uniformly remesh the input mesh and compute the paths applying the well known Dijkstra shortest path algorithm \cite{Dijkstra1959}. Additionally, the constraints recomputation illustrated in Fig. \ref{fig:subcontours} acts as a correction factor reducing the influence of the triangulation and the specific seed points used on the definition of the boundaries. According to our experiments this procedure is sufficient to define reproducible boundaries.
However, if high accuracy is required and to have higher independence on the mesh, other methods could be used to compute more accurate geodesics \cite{Surazhsky2005,Crane2013}.

Side-by-side or overlaying inspection of different FRF maps is convenient in many clinical situations: to analyse the temporal evolution of some parameter (e.g. pre- and post-ablation fibrosis extent) from the same patient; to compare and correlate different features from the same patient (e.g. gadolinium enhancement from LGE-CMR data and endocardial voltage from electroanatomical maps); to compare same feature in different patients (e.g. regional or global fibrosis extent \cite{benito2018preferential}), etc. We have shown pairwise comparisons of LGE-CMR intensities and bipolar endocardial voltage but a better depiction of the electroanatomical maps is required (e.g. using high definition mapping systems like Rhythmia (HDx Mapping System, Boston Scientific)) to achieve realistic representations of the LA and be able to fairly compare it with other kinds of data such as LGE-MRI.

Proposing a standard 2D template is far from trivial. We observed how depending on the specific morphology of the 3D LA different configurations (radius and position of the holes) were preferable. We also observed that quasi-conformal mappings (without any additional constraint) do not typically provide intuitive and clinically useful LA representations. Methods aiming at optimizing the quasi-conformality of the transformation \cite{Ho2016} generate 2D depictions where the surroundings of the PVs (the most clinically relevant region) are mapped to a small central area of the disk and appear therefore very squeezed (see Fig. 3a). The method proposed in \cite{Ho2016} could have been used to define a template but we believe that it will generate very different templates depending on the specific 3D mesh used. One 3D LA should therefore be used as a reference (posing the problem that the template will not be optimal for the most morphologically different LAs) or several 3D or 2D templates must be averaged somehow. In our approach, we prioritized the interpretability of the 2D template, which must be intuitive, a representation of the LA  that clinicians like and are used to see. The proposed template
can be used for most of the LA shapes (with 4 veins), which takes into account the real averaged inter-vein (and LAA) separation and facilitates visualization and interpretability of the 2D map. 
However, the estimated metrics (proportions) were obtained using a dataset of AF patients, i.e. bigger and more spherical LA, and with bigger PV ostia than controls. \cite{wozniak2011comparison} showed how diameters of PV ostia were larger in AF patients than in control subjects in accordance with increased LA volume of the AF patients. 
Nonetheless, we believe that the changes in the size of the atrium or ostia will not change significantly our distortion analysis. 

The method presents some limitations. First, it can only be applied to LA with 4 PVs. Other templates can be derived similarly to the one proposed here but direct comparison with the 4-veins template will not be accurate in regions close to the PV holes. The method can be applied to atria with a common left trunk if there exist a distant bifurcation but it should be taken into account that the region represented between the veins would not be the carina but the common left trunk itself. LA with extra veins (e.g. extra middle right PV) can also be flattened with our method if the vein can be clipped or ignored. This solution may be acceptable depending on the application. For example, it could be done to quantify global extent of fibrosis but it should be avoided if the representation of the surroundings of the PVs is important, for example in gap detection after pulmonary vein isolation.
Our method may provide unsatisfactory results in the case of low resolution meshes (few points) and with PVs very close to each other. In that case, the number of points between the veins may not be sufficient to depict the template and the triangles may appear stretched. This problem can be typically solved by increasing the number of points remeshing the given low resolution mesh \cite{valette2008}. 
We set the weighting factor, $w$, to 1000 because it was empirically found to work well in all the cases. This factor enforces that the \emph{boundary} points (MV, PVs and LAA) are correctly placed in the target positions and the conformality of the inner points over the fulfillment of the \emph{regional constraints} (that can be therefore considered as soft constraints). 


\section{Conclusions}
We have presented a method to unfold any 3D LA surface and depict it in an intuitive, standardised, two-dimensional map. Compared to the SUM it does not rely on a 3D registration step, much faster and guarantees no information loss, as it could happen due to a bad registration or information interpolation when the source and target meshes have very different shapes and resolutions. 
Flattening and registration is achieved simultaneously and regional interpretability is emphasized by confining anatomical regions to predefined regions in the 2D map avoiding undesired displacements attributable to global flattening methods.
We have shown the suitability of our method to easily inspect the whole LA anatomy in 1 view, and to consistently divide the different LA anatomies and put them into correspondence making its regional comparison possible.     
The implementation is interactive, computationally efficient, fast, and easy to use. The code (Python) is publicly available\footnote{\url{https://github.com/martanunez/LA_flattening}} and it can be run in a normal PC without needing high computational resources or advanced technical skills.

\ifCLASSOPTIONcompsoc
  \section*{Acknowledgments}
\else
  \section*{Acknowledgment}
\fi

This study was partially funded by the Spanish Ministry of Economy and Competitiveness (DPI2015-71640-R), by the ``Fundaci\'o La Marat\'o de TV3" (n\textsuperscript{o} 20154031) and by European Union Horizon 2020 Programme for Research and Innovation, under grant agreement No. 642676 (CardioFunXion).

\ifCLASSOPTIONcaptionsoff
  \newpage
\fi

\vspace{-1cm}

\begin{IEEEbiographynophoto}{Marta Nu\~nez-Garcia}
is a postdoctoral researcher at IHU Liryc (L'Institut de Rythmologie et Modelisation Cardiaque, Bordeaux, France). She received her PhD in Medical Imaging from the Universitat Pompeu Fabra (Barcelona, Spain) in 2018. 
\end{IEEEbiographynophoto}
\vspace{-1.5cm}

\begin{IEEEbiographynophoto}{Gabriel Bernardino} is a PhD candidate at Universitat Pompeu Fabra, at Information and Communication Technologies Department. 
Bernardino received a  MSc in Mathematics from Rheinische Friedrich-Wilhelms Universit\"{a}t Bonn.
\end{IEEEbiographynophoto}
\vspace{-1.5cm}

\begin{IEEEbiographynophoto}{Francisco Alarc\'on} is a predoctoral researcher at the Arrhythmia Section, Hospital Cl\'inic Barcelona. He received his MSc in Biomedical Engineering from the Universitat de Barcelona in 2016. 
\end{IEEEbiographynophoto}
\vspace{-1.5cm}

\begin{IEEEbiographynophoto}{Gala Caixal} is a predoctoral researcher at Universitat de Barcelona (UB) and the Institut d'Investigacions Biom\`ediques August Pi i Sunyer (IDIBABS). She received her degree in Medicine from the UB and her specialization in Cardiology from the Hospital Cl\'inic. 
\end{IEEEbiographynophoto}
\vspace{-1.5cm}

\begin{IEEEbiographynophoto}{Dr. Llu\'is Mont} MD, PhD, FESC, FEHRA, is Professor of Medical School at Universitat de Barcelona, head of the Arrhythmia Section, and responsible of the Atrial Fibrillation Unit of the Hospital Cl\'inic de Barcelona. Dr. Mont graduated in Medicine and Surgery in 1981 by the School of Medicine of Universitat de Barcelona. He obtained his Cardiologist Certificate in 1988. 
He has been involved in several research projects and multicenter studies about Atrial Fibrillation and Resyncrhonization Therapy.
\end{IEEEbiographynophoto}
\vspace{-1.5cm}

\begin{IEEEbiographynophoto}{Oscar Camara} received 
the MSc and PhD degrees in image processing from the \'Ecole Nationale Sup\'erieure des T\'el\'ecommunications, Paris, in 2000 and 2003, respectively. 
In 2007, he joined Universitat Pompeu Fabra (UPF) as a Ram\'on y Cajal Fellow and later became an Associate Professor in 2012. He is coordinating the PhySense Research Group, which he founded in 2011. He is also one of the founders of the BCN-Medtech unit. 
\end{IEEEbiographynophoto}
\vspace{-1.5cm}

\begin{IEEEbiographynophoto}{Constantine Butakoff} is a postdoctoral researcher at Barcelona Supercomputing Center. 
He received a PhD in biomedicine in 2009 from I3A (Instituto de Investigacion en Ingenieria de Aragon), Universidad de Zaragoza, Spain. 
\end{IEEEbiographynophoto}








\clearpage

\appendices
\section{Data description.}

A detailed description of the different datasets used in this paper is provided next:

\begin{enumerate}

\item  \textbf{Training set from the 2018 Atrial Segmentation Challenge (STACOM 2018)\footnote{\url{http://atriaseg2018.cardiacatlas.org/}}}. 
Expert manual segmentations of the endocardium were acquired from LGE-CMR data with resolution 0.625 x 0.625 x 1.25 mm$^3$. The well known marching cubes (MC) algorithm was then applied and given the high resolution of the images the result was a smooth 3D surface. We used the MC implementation available in the VTK library. We then uniformly remeshed the meshes using the methodology described in \cite{valette2008} and the implementation provided in \url{https://github.com/valette/ACVD}. The target number of points was set to 20000. This dataset was used to define the 2D template (Figs. \ref{fig:boxplots} and \ref{fig:dist_metrics}) and to compute an averaged LA mesh using Deformetrica \cite{bone2018deformetrica}. Examples of this dataset can be seen in Fig. \ref{fig:example} where voxel intensities from LGE-CMR images were mapped onto the LA surface mesh using the maximum intensity projection (MIP) technique: images were sampled along the normals on both sides of the surface mesh, assigning to the corresponding vertex the maximum intensity value sampled. The depth of the sampling was set to 3 mm which is considered to be large enough to overcome potential segmentation errors while at the same time sufficiently small to avoid going too far from the atrial wall \cite{nunez2019mind}.

\item {\textbf{Averaged LA mesh.}}
The mean LA mesh shown in Figs. \ref{fig:meshes_flat}, \ref{fig:circles} and \ref{fig:pattern_appen}, was obtained from the meshes described in the previous point. We first extracted the LA cavities using the methodology explained in Section \ref{subsec:step1} (step 1 in Fig. \ref{fig:pipeline}); the cavities were then rigidly aligned using the iterative closest point (ICP) algorithm; and finally the averaged mesh was computed using the Deformetrica’s atlas construction tool \cite{bone2018deformetrica}. As a result, the averaged mesh was also smooth and the triangulation roughly uniform. We anyways applied uniform remesh \cite{valette2008} to obtain the final averaged mesh (15000 points).

\item \textbf{SUM template \cite{williams2017standardized}.} 
As explained by the authors, this template was obtained from an average atlas image but any further information is provided. The number of points, 6132, is rather low.

\item \textbf{CARTO (Biosense Webster Inc, Diamond Bar, CA)\footnote{\url{https://www.biosensewebster.com/products/carto-3.aspx}} exported meshes.} The CARTO electroanatomical mapping system reconstructs LA shapes from a cloud of points acquired with an intracavitary catheter. The exported maps are already smooth and with relatively high number of points (more than 20000) but we routinely apply a uniform remesh (followed by information projection) before the flattening. The examples shown in this manuscript (Figs. \ref{fig:gaps}.2 left and \ref{fig:eam_reso} left) were obtained in clinical routine at the Hospital Clinic (Barcelona, Spain) and all patients signed written informed consent.

\item \textbf{Manual segmentations from LGE-CMR images.}
We also used the set of LGE-CMR images corresponding to the same patients in the previous point. Image resolution was 1.25 x 1.25 x 2.5 mm$^3$ and LA segmentations were obtained using the ADAS-AF software (Galgo Medical S.L., Barcelona, Spain)\footnote{\url{https://www.adas3d.com/en/adas-af-product.html}}. Expert manual delineation of the mid-wall of the LA was performed in each transverse LGE-CMR slice plane. The software then generated a 3D smooth surface of the LA from the slice-by-slice segmentation. Finally, image intensity from the LGE-CMR data was projected onto the surface. Figs. \ref{fig:gaps}.1, \ref{fig:gaps}.2 right and \ref{fig:eam_reso} right show examples of these dataset. 

\end{enumerate}

\section{Shape metrics.}

\begin{table*}[h!]\footnotesize
\caption{Mean, standard deviation (SD) and median of the PVs and LAA ostium perimeters (left) and perimeter ratio with respect to LIPV which is the smallest one and considered as reference (right). All distances are reported in mm. LIPV = left inferior PV; LSPV = left superior PV; RIPV = right inferior PV; RSPV = right superior PV; LAA = left atrial appendage.}
\centering
\begin{tabular}{cccccc|cccc}
& \multicolumn{5}{c|}{\textbf{Perimeter (mm)}} & \multicolumn{4}{c}{ \textbf{Perimeters ratio}} \\ 
\hline
    &  \textbf{LIPV} & \textbf{LSPV}& \textbf{RIPV}& \textbf{RSPV} & \textbf{LAA} & \textbf{$\frac{LSPV}{LIPV}$} & \textbf{$\frac{RIPV}{LIPV}$} & \textbf{$\frac{RSPV}{LIPV}$} & \textbf{$\frac{LAA}{LIPV}$}\\
   \hline
    \textbf{Mean} & 38.73 & 43.06 & 44.49 & 50.47 &  54.30 & 1.17 & 1.18 & 1.36 & 1.48\\
    \textbf{SD} & 9.17 & 7.37 & 9.91 & 11.21 & 14.06 & 0.35 & 0.29 & 0.39 & 0.50\\
   \textbf{Median} & 38.32 & 41.84 & 46.11 & 49.39  & 52.61 & 1.14 & 1.13 & 1.34 & 1.39\\
\end{tabular}%
\label{tab:perimeters}
\end{table*}

\begin{table*}[h]
\caption{Mean, standard deviation and median of the inter-veins and LAA distance (left) and distance ratio with respect to the LSPV-LIPV (left carina) distance which is considered as reference (right). All distances are reported in mm. LIPV = left inferior PV; LSPV = left superior PV; RIPV = right inferior PV; RSPV = right superior PV; LAA = left atrial appendage.}
\centering
\resizebox{\textwidth}{!}{%
\begin{tabular}{cccccc|cccc}
& \multicolumn{5}{c|}{\textbf{Distance (mm)}} & \multicolumn{4}{c}{ \textbf{Distance ratio}} \\ 
\hline
    &  \textbf{LSPV-LIPV} & \textbf{RSPV-RIPV}& \textbf{RSPV-LSPV}& \textbf{RIPV-LIPV} & \textbf{LSPV-LAA} & \textbf{$\frac{RSPV-RIPV}{LSPV-LIPV}$} & 
\textbf{$\frac{LSPV-LAA}{LSPV-LIPV}$} & \textbf{$\frac{RSPV-LSPV}{LSPV-LIPV}$} & \textbf{$\frac{RIPV-LIPV}{LSPV-LIPV}$}\\
\hline   
    \textbf{Mean} & 14.82 & 17.03 & 57.66 & 58.91 &  23.94 & 1.18 & 1.72 & 4.06 & 4.15\\
    \textbf{SD} & 3.29 & 5.62 & 8.61 & 7.21 & 7.95 & 0.36 & 0.75 & 1.01 & 0.99 \\
   \textbf{Median} & 13.96 & 15.88 & 57.98 & 60.29  & 22.54 & 1.09 & 1.59 & 3.72 & 3.81\\
\end{tabular}%
}
\label{tab:distances}
\end{table*}


\section{Additional flattening examples.}
\renewcommand{\thefigure}{C.\arabic{figure}}
\setcounter{figure}{0}

\begin{figure*}[h]
\centering
\includegraphics[width=0.8\textwidth]{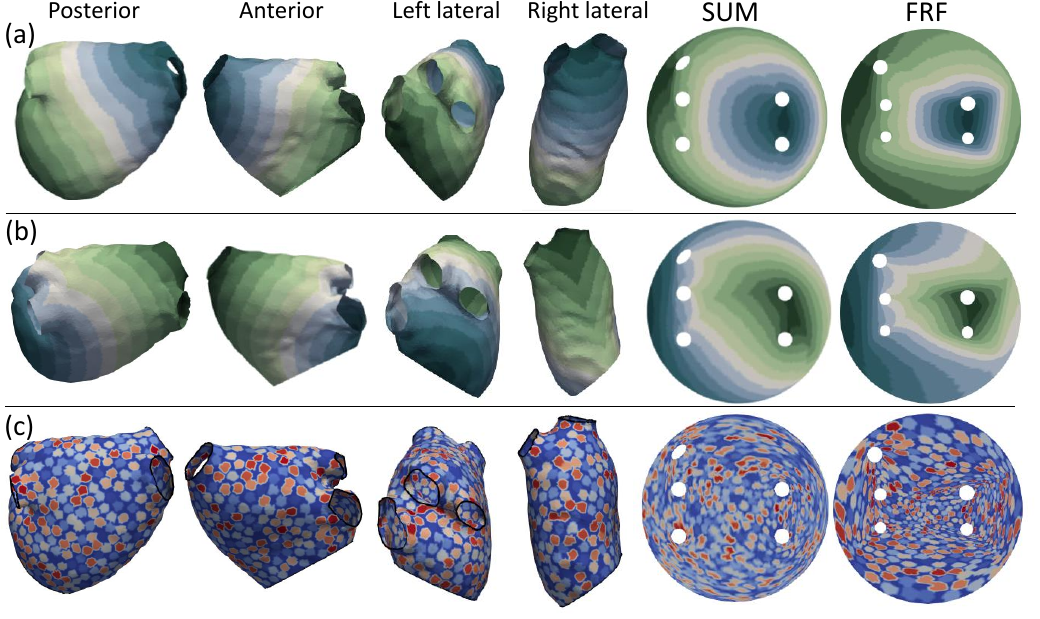} 
\caption[Additional flattening examples of 3D LA with different synthetic textures on the surface.]{Additional flattening examples of 3D LA with different synthetic textures on the surface and corresponding SUM and FRF maps. First and second rows show striped synthetic patterns, defined on the 3D surface representing isocontours of the distance to a seed point placed in the right carina (a), and the MV contour (b). (c) shows a texture of 500 uniformly distributed spots.}
\label{fig:pattern_appen}
\end{figure*}

\begin{figure*}[th]
\centering
\includegraphics[width=0.80\textwidth]{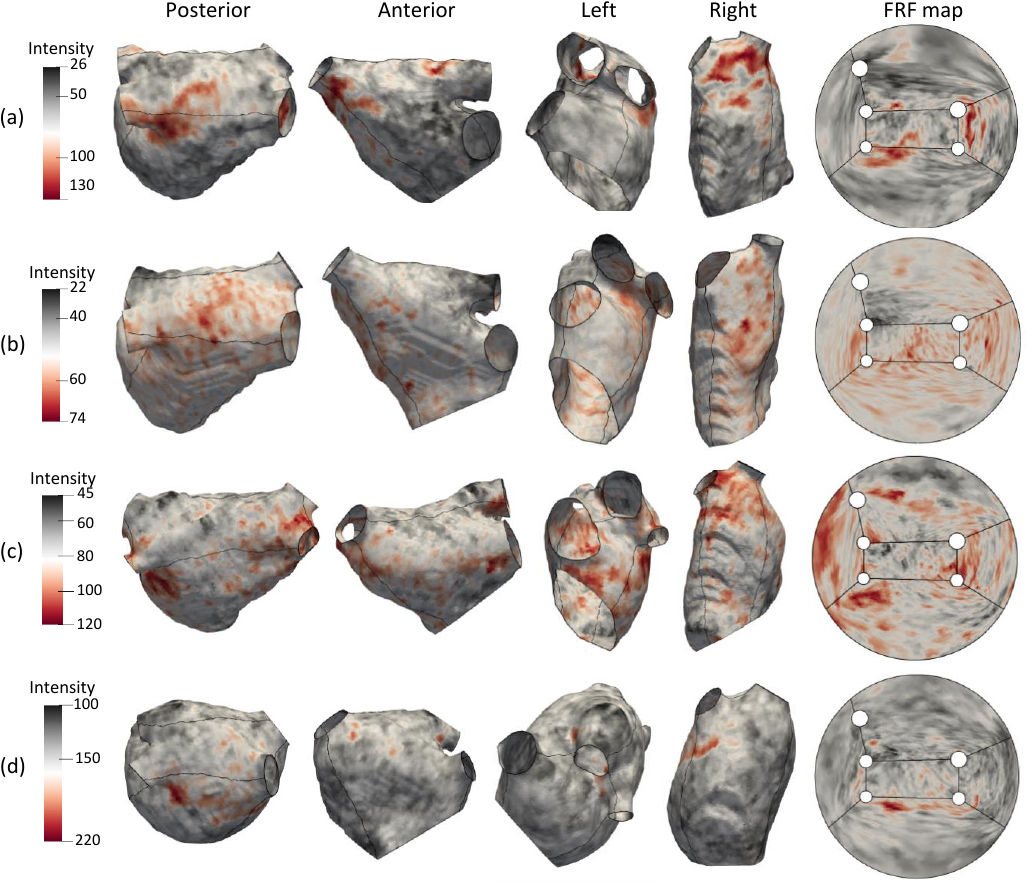} 
\caption[Examples of flattened LA with projected LGE-CMR intensity.]{Examples of LA flattening showing 4 views of manually segmented LA surfaces with projected late gadolinium enhanced magnetic resonance imaging (LGE-CMR) intensities and corresponding FRF maps.}
\label{fig:example}
\end{figure*}

\begin{figure*}[t]
\includegraphics[width=\textwidth]{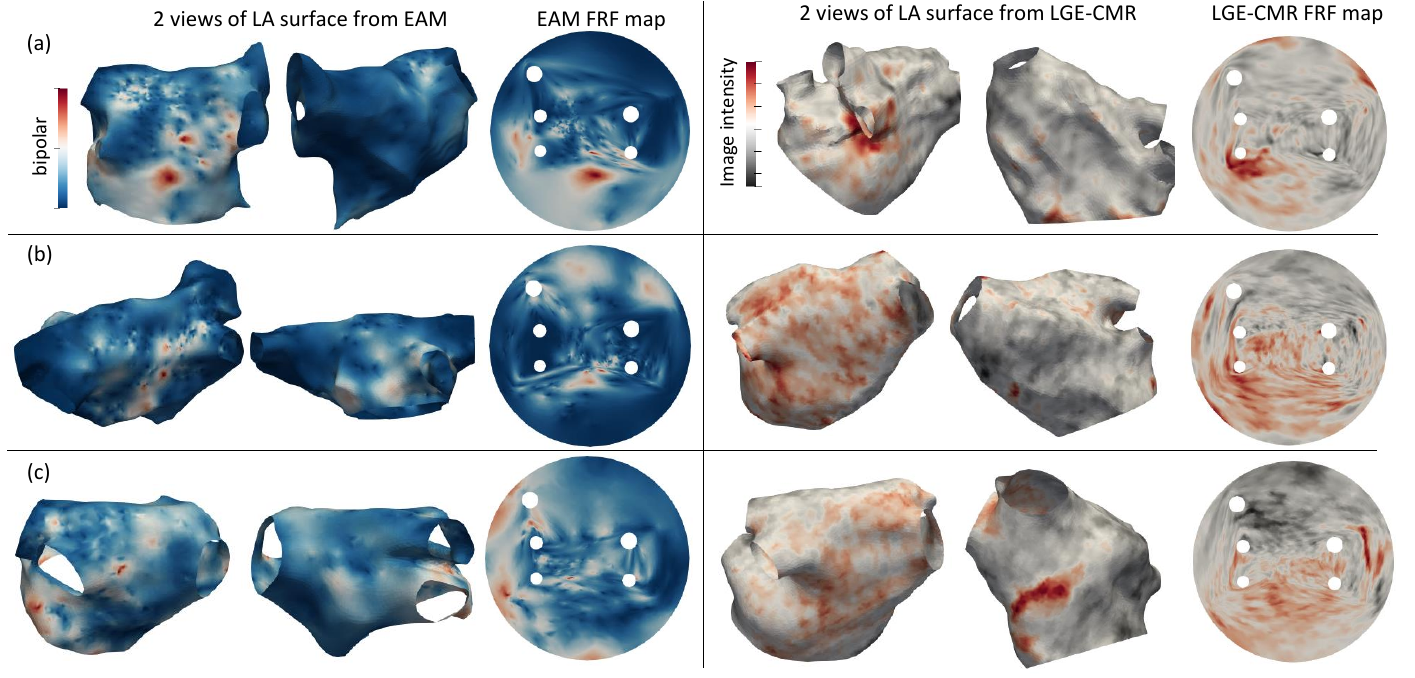}
\caption[FRF maps corresponding to EAM and LA segmentations from LGE-CMR data.]{Three examples of EAM and corresponding (same patient) LA from LGE-CMR with projected image intensity. Low voltage (dark blue) is typically associated to high LGE-CMR intensity (red).}
\label{fig:eam_reso}
\end{figure*}

\end{document}